# The Elemental Abundances of Ryugu: Assessment of Chemical Heterogeneities and the Nugget Effect

Tetsuya Yokoyama[1*], Nicolas Dauphas[2], Ryota Fukai[3], Tomohiro Usui[3], Shogo Tachibana[4], Maria Schönbächler[5], Henner Busemann[5], Masanao Abe[3], and Toru Yada[3]

[1] Department of Earth & Planetary Sciences, Institute of Science Tokyo, Japan
[2] Origins Lab, Department. of the Geophysical Sciences and Enrico Fermi Institute, The University of Chicago, USA
[3] Astromaterials Science Research Group, ISAS, JAXA
[4] Department of Earth and Planetary Science, University of Tokyo, Japan
[5] Institut für Geochemie und Petrologie, ETH Zürich, Switzerland

*Corresponding author E-mail: tetsuya.yoko@eps.sci.titech.ac.jp

## Abstract

The Hayabusa 2 spacecraft sampled ~5.4 g of asteroid material from the Cb-type asteroid Ryugu. Initial analysis of the Ryugu materials revealed a mineralogical, chemical, and isotopic kinship to the CI chondrites. In this study, we have summarized the elemental abundances of Ryugu samples published to date, and evaluated their compositional variability associated with the CI chondrite data. The abundances of some elements (e.g., P, Ca, Mn, and rare earth elements) in individual Ryugu particles were found to show large relative dispersions compared to the other elements, presumably due to the nugget effect of aqueously formed minor secondary minerals (e.g., dolomite, apatite, magnetite, and pyrrhotite). Consequently, the mean abundances of Ryugu for these elements, calculated using currently available Ryugu data, are accompanied by a certain degree of uncertainties. We suggest establishing a consortium to determine the representative elemental abundances of Ryugu by measuring aliquots from a large homogenized powder sample that can mitigate the nugget effect. Our statistical calculation shows that at least 750 and 400 mg of homogenized samples from Chambers A and C, respectively, are needed to achieve within ±5% compositional heterogeneity. The data obtained throughout the consortium activity complement the scientific objectives of the Hayabusa2 mission. Moreover, we anticipate that the obtained Ryugu data, coupled with the elemental abundances of CI chondrites, provide new insights into the chemical composition of the Solar System, which will be used by multidisciplinary communities, including Earth and planetary sciences, astronomy, physics, and chemistry.

## 1. Introduction

Multidisciplinary studies in astronomy, planetary science, and cosmochemistry have revealed that the Solar System began ~4.6 billion years ago, triggered by the gravitational collapse of a molecular cloud core to which various stellar objects that existed before the birth of the Solar System contributed (e.g., Amelin *et al.*, 2002; Cameron, 1988; Connelly *et al.*, 2012). Accurate determination of the initial composition of the Solar System is critical for advancing our understanding of the formation and evolution of the Solar System objects including planets, moons, asteroids, and comets. Additionally, the chemical composition of the Solar System serves as an important baseline for studying the chemical composition of other stars and exoplanets, as well as the interstellar medium and galaxies.

The elemental abundances of the Sun have been used to represent the chemical composition of the Solar System, since more than 99% of the mass of the Solar System is locked up in the Sun. There are several ways to determine the elemental abundances of the Sun, and each method provides complementary information about the composition of the Sun. One of the most important and direct methods for determining the elemental abundances of the Sun is spectroscopy, which involves the analysis of dark lines at specific wavelengths corresponding to the absorption of light by atoms in the solar photosphere. Spectroscopy of the solar photosphere in determining the elemental abundances of the Sun has several difficulties, including mixing of multiple spectral lines, uncertainties in the solar atmospheric models that describe the physical conditions of the Sun, the stratification of elemental abundances in the upper and lower layers of the solar atmosphere, and the analysis of weak lines for extremely low-abundance elements. Despite these difficulties, the abundances of 68 elements in the solar photosphere are reported with uncertainties of 10–20% for about two-thirds of these elements, whereas the remainder have uncertainties >20%, specifically for minor elements and volatiles (Lodders, 2021).

Another robust method for determining the chemical composition of the Solar System is the measurement of chondritic meteorites, which are fragments of asteroids that did not melt after accretion in the early Solar System (Scott and Krot, 2014). Of all the chondrites, which account for about 90% of the >70,000 meteorites discovered on Earth to date (Meteoritical Bulletin Database: https://www.lpi.usra.edu/meteor/metbull.php), the Ivuna-type (CI) carbonaceous chondrites have been perceived as a unique group of meteorites with a chemical composition similar to that of the solar photosphere except for highly volatile elements (noble gases, H, C, N, and O) and Li that was destroyed in the Sun by nuclear reactions. In fact, direct comparisons of relative elemental abundances between solar photosphere and CI chondrites showed excellent agreement within ±10% difference for nearly 40 elements, regardless of their volatility and geochemical character (Lodders *et al.*, 2009; Palme *et al.*, 2014). Although these authors found several problematic elements exceeding the ±20% difference limit (e.g., Mn, Ga, Rb, Hf, and W), such a situation would be mitigated by further improvements in the spectroscopic analyses of the solar photosphere, especially by the use of 3D solar atmospheric models (e.g., Asplund *et al.*, 2021; Magg *et al.*, 2022). It should be noted that CI chondrites are extremely rare, only five witnessed fall CIs have been collected so far (Alais, Ivuna, Orgueil, Revelstoke, and Tonk). A recent study argued that the elemental abundances of Ivuna, Alais, and Tonk agree well with the results of the Orgueil analyses, with only a few exceptions (Palme and Zipfel, 2021). This observation led the authors to conclude that all CI chondrites have essentially the same fractions of the fundamental cosmochemical components, and that the CI composition is a well-defined entity representing the non-gaseous compositions of the solar nebula and photosphere of the Sun.

An additional advantage of CI chondrite measurements is that not only elemental abundances but also isotopic compositions can be measured directly in the laboratory with high precision (at the ppm level). Solar wind measurements on samples collected by the Genesis mission allow a comparison of oxygen isotopic compositions between CI chondrites and the Sun (McKeegan *et al.*, 2011), although it should be noted that solar wind isotopic compositions may not be representative of the Solar System as a whole, considering the influence of mass-dependent isotopic fractionation (Lodders, 2021). On the other hand, a severe problem associated with the measurement of meteorites is terrestrial weathering, which changes the original mineralogical and chemical composition (Bland *et al.*, 2006). This is especially true for find meteorites that are recovered after a certain residence time at specific environments on Earth (e.g., Antarctica and hot deserts), although the compositional change due to terrestrial weathering is also unavoidable for fall meteorites. Compositional changes caused by terrestrial weathering include oxidation of Fe in metal and sulfides, hydration/hydrolysis of silicates, mobilization of S, dissolution and loss/gain of fluid mobile elements (e.g., Na, Sr, Ba, and U), and degradation of organic C (e.g., Bland *et al.*, 2006; Friedrich *et al.*, 2002; King *et al.*, 2020). Soils can contaminate meteorites when a meteor enters the Earth and hits the ground. Additionally, white sulfate forms on the surface and in the interior veins of CI chondrites during their long-term storage in museums (Gounelle and Zolensky, 2001; King *et al.*, 2020).

Sample return missions are superior to meteorite analysis in that samples can be collected with no or a minimum of contamination from well-documented extraterrestrial objects. The Japan Aerospace Exploration Agency's (JAXA) Hayabusa 2 spacecraft, targeting the Cb-type asteroid (162173) Ryugu, sampled ~5.4 g of asteroidal material and returned the samples to Earth in December 2020 (Tachibana *et al.*, 2022; Yada *et al.*, 2022). These samples were collected during the two landing sequences on the asteroid Ryugu. During the first touch-down operation (TD1), samples were collected from the asteroid surface and stored in sample Chamber A, while the other samples stored in sample Chamber C were collected from the vicinity of an artificial crater created by the small carry-on impactor during the second touch-down operation (TD2). The TD1 and TD2 samples were stored and handled separately at the JAXA curation facility (Yada *et al*. 2022).

Initial analyses of the Ryugu materials in both chambers revealed a mineralogical and chemical kinship to the CI chondrites (Nakamura *et al.*, 2022; Nakamura *et al.*, 2023; Yokoyama *et al.*, 2023a) with a composition similar to the solar photosphere except for highly volatile elements (Lodders, 2021). Among the non-volatile elements, Ta stands out with a large excess in those Ryugu samples sampled during the second touchdown, but this is due to contamination from the Ta projectile used for sampling from larger depth of Ryugu (Nakamura *et al.*, 2022; Yokoyama *et al.*, 2023a). Isotopic analyses of Ryugu materials showed that Ryugu and CI chondrites presumably originated from the outskirts of the Solar System (Hopp *et al.*, 2022; Kawasaki *et al.*, 2022; Paquet *et al.*, 2023).

The present study summarizes the elemental abundances of Ryugu bulk samples published to date, evaluates the compositional variability, and compares the results with those of CI chondrites. In particular, the influence of the heterogeneous distribution of minor secondary minerals formed during aqueous alteration of the parent body is discussed to assess the dispersion of elemental abundances among different Ryugu particles. We then demonstrate the scientific need for a consortium to determine the representative elemental abundances of Ryugu using a relatively large amount of homogenized powder sample. The data obtained throughout the activity of the consortium should be used to complement the scientific objectives of the Hayabusa2 mission, including investigations of (1) the evolution from a planetesimal to a near-Earth asteroid, (2) the possible destruction and accumulation of a rubble-pile body formed from a larger, aqueously altered parent planetesimal, (3) the diversification of organic materials through interactions with minerals and water in a planetesimal, and (4) the chemical heterogeneity in the early Solar System (Tachibana *et al.*, 2014). Furthermore, we anticipate that the obtained Ryugu data, coupled with the elemental abundances of CI chondrites, provide new insights into the chemical composition of the Solar System, which will be beneficial for multidisciplinary communities in various scientific fields, extending beyond the scope of the Hayabusa2 project.

## 2. Summary of the reported Ryugu data

### 2.1. Elemental abundances of bulk Ryugu samples

Six previous studies have measured elemental abundances of bulk Ryugu samples using different approaches. Yokoyama *et al*. (2023a) quantified the abundances of 66 elements in the Ryugu samples by X-ray fluorescence (XRF), inductively coupled plasma mass spectrometry (ICP-MS), thermogravimetric analysis coupled with mass spectrometry (TG-MS), and combined analyses of pyrolysis and combustion. In the XRF analysis, 33 mg of an aggregate Ryugu sample C0108 was first powdered, of which 24 mg was set in an acrylic sample cell to determine 22, 17, and 18 elements by wavelength dispersive (WD), energy dispersive (ED), and high energy (HE) XRF, respectively. The powdered sample C0108 in the cell was retrieved after the XRF measurements, and acid digested to a homogeneous solution for ICP-MS measurements of 54 elements. Additionally, 24 mg of a powdered sample A0106-A0107, prepared from a mixed aggregate of A0106 (1.6 mg) and A0107 (27 mg), was acid digested into a homogeneous solution for ICP-MS analysis. Therefore, these elemental abundances represent the bulk chemical composition in a relatively large size (24 mg) of Ryugu samples. On the other hand, two ~1 mg aliquots of A0040 were used in the TG-MG and pyrolysis measurements to determine the abundances of H and C.

Nakamura *et al*. (2022) examined 16 individual Ryugu particles (A0022, A0033, A0035, A0048, A0073, A0078, A0085, C0008, C0019, C0027, C0039, C0047, C0053, C0079, C0081, and C0082) for comprehensive geochemical analyses, ranging in size from 1.2 to 3.7 mm (largest dimension) and weight from 0.7 to 10.0 mg. For the determination of elemental abundances, an aliquot from each particle (0.33–3.3 mg) was separated and further divided into up to 8 portions to determine the abundances of different groups of elements. Using these aliquots, 66 elements were measured by ICP-MS, while the abundances of Ne and H, C, and N were determined by noble gas mass spectrometry (noble gas MS) and isotope ratio mass spectrometry (IR-MS), respectively. The abundance of Si was not measured directly, but calculated from the Si/Mg ratio obtained by SEM-EDS elemental mapping of the entire particle cross-section prepared by an ultramicrotome, and the Mg abundance determined by ICP-MS analysis. In contrast to Yokoyama *et al*. (2023a), the elemental abundances determined by Nakamura *et al*. (2022) are determined in separated small aliquots and not powdered, and thus do not necessarily represent the bulk chemical composition of individual particles.

Ito *et al*. (2022) measured the abundances of 20 elements in two Ryugu particles (1.06 mg of A0098; 0.65 mg of C0068) by instrumental neutron activation analysis (INAA). Okazaki *et al*. (2023) determined noble gas, C, and N abundances in two Ryugu samples (A0105 and C0106) by mass spectrometry. Naraoka *et al*. (2023) and

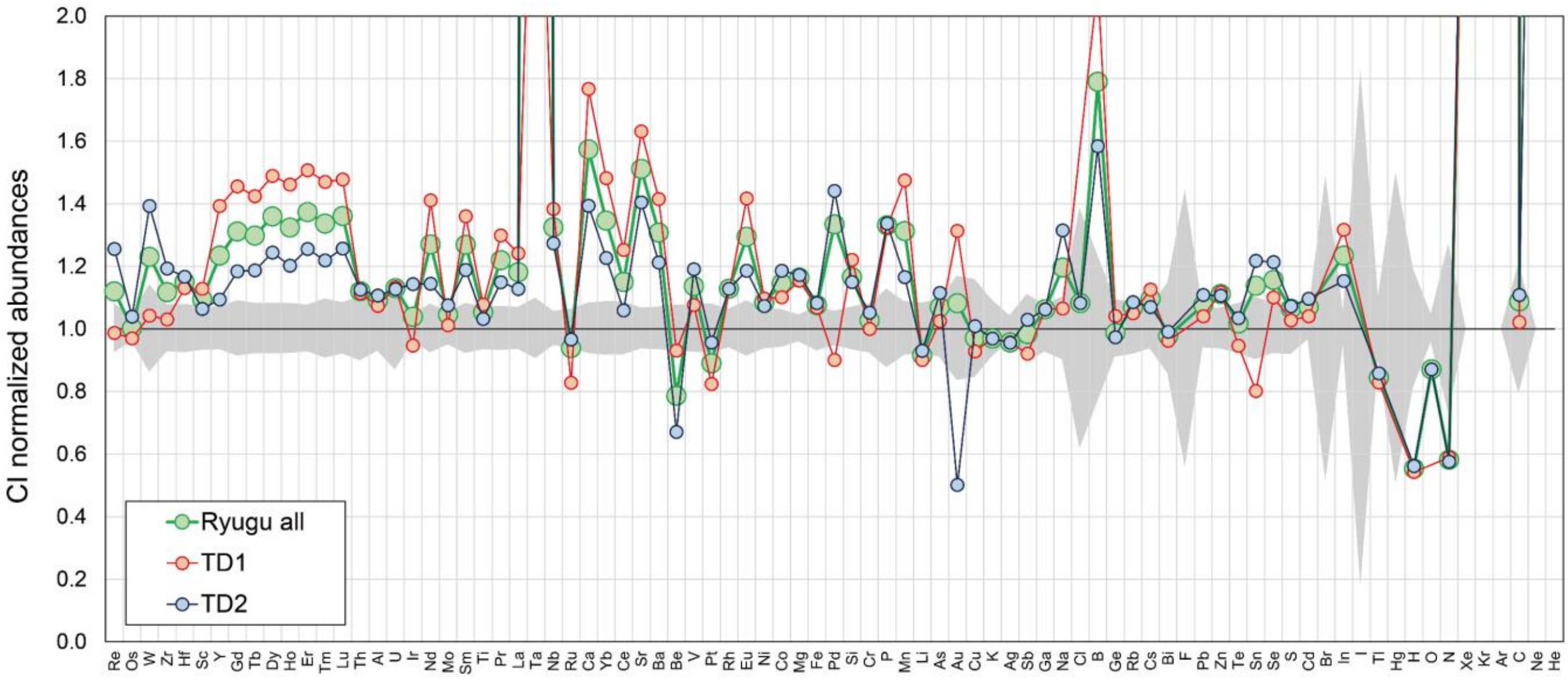

**Fig. 1** CI-normalized elemental abundances in bulk Ryugu samples from (a) Chambers A ("TD1") and C ("TD2") combined, (b) Chamber A, and (c) Chamber C determined by six previous studies (Ito et al., 2022; Nakamura et al., 2022; Naraoka et al., 2023; Oba et al., 2023; Okazaki et al., 2023; Yokoyama et al., 2023a). CI chondrite data and their 2-sigma standard deviations (±2σ; gray band) were taken from Lodders (2021). Elements are presented in the order of decreasing 50% condensation temperature of Lodders (2003) from left to right. No further normalization to an element (e.g., Mg, Si, Fe) was performed.

Oba *et al*. (2023) measured the abundances of H, C, N, and S in Ryugu samples A0106 and C0107 by elemental analyzer (EA)-IRMS, respectively. In total, the six previous studies mentioned above measured the abundances of 79 elements (**Table S1**). Lodders (2021) reported the abundances of 83 elements as representative for CI chondrites. The elements that have not yet been measured for bulk Ryugu samples are F, Br, I, and Hg.

In the six previous studies, different methods were applied to different batches of samples, some of which were very small (e.g., <1 mg) and thus potentially chemically heterogenous (see below). This makes it difficult to accurately determine "bulk" Ryugu elemental abundances using established statistical protocols performed by previous compilation studies on reference materials (e.g., Jochum *et al.*, 2016). Here, we present the current best estimates for the Ryugu elemental abundances with a straightforward approach that calculates the averages of the reported data, taking into account the measured sample weights;

$$C = \frac{\sum C_i m_i}{\sum m_i} \qquad (1)$$

where $C_i$ and $m_i$ are the mass fraction and weight, respectively, of the $i$-th measurement of a target element. **Table 1** summarizes the results of the estimated bulk Ryugu abundances for 79 elements in TD1, TD2, and all samples. The uncertainties $\sigma_C$ of the estimated abundances were calculated by propagating the uncertainties associated with $C(X)_i$ and $m_i$ using

$$\sigma_C = C\sqrt{\frac{\sum C_i^2\sigma_{mi}^2 + \sum m_i^2\sigma_{Ci}^2}{(\sum C_i m_i)^2} + \frac{\sum \sigma_{mi}^2}{(\sum m_i)^2}} \qquad (2).$$

**Figure 1** shows the estimated elemental abundances of bulk Ryugu samples normalized to CI chondrite values (Lodders, 2021). As discussed in previous studies, the Ryugu elemental abundances show close agreement with CI chondrites from refractory to volatile elements, with slight excesses in most elements (Nakamura *et al.*, 2022; Yokoyama *et al.*, 2023a). The observed difference in the elemental abundances between Ryugu and CI chondrites will be discussed in the following sections.

## 2.2. Isotopic abundances of bulk Ryugu samples

The above six previous studies measured the bulk isotopic compositions of various elements in the Ryugu samples used in the measurement of bulk elemental abundances (Ito *et al.*, 2022; Nakamura *et al.*, 2022; Naraoka *et al.*, 2023; Oba *et al.*, 2023; Okazaki *et al.*, 2023; Yokoyama *et al.*, 2023a). These include H, C, N, O, S, Ca, Ti, Cr, and noble gas elements (He, Ne, Ar, Kr, and Xe). Additional measurements of bulk isotopic compositions were performed using the same sample aliquots described in Yokoyama *et al*. (2023a) for Mg

**Table 1** Elemental abundances of bulk Ryugu samples

| *Z* | *E* | Ryugu (TD1) | | | | | Ryugu (TD2) | | | | | Ryugu (All) | | | | | CI (Lodders 2021) | | |
|---|---|---|---|---|---|---|---|---|---|---|---|---|---|---|---|---|---|---|---|
| | | µg/g | | 2σ | n | mass (mg) | µg/g | | 2σ | n | mass (mg) | µg/g | | 2σ | n | mass (mg) | µg/g | | 2σ |
| 1 | H | 10098 | ± | 115 | 8 | 11.9 | 10443 | ± | 121 | 8 | 15.1 | 10291 | ± | 85 | 16 | 27.1 | 18600 | ± | 3440 |
| 2 | He | 0.189 | ± | 0.006 | 12 | 0.903 | 0.0252 | ± | 0.0010 | 5 | 0.240 | 0.155 | ± | 0.005 | 17 | 1.14 | 0.00917 | | |
| 3 | Li | 1.36 | ± | 0.09 | 8 | 34.6 | 1.40 | ± | 0.07 | 10 | 39.1 | 1.38 | ± | 0.06 | 18 | 73.6 | 1.51 | ± | 0.12 |
| 4 | Be | 0.0205 | ± | 0.0052 | 6 | 30.9 | 0.0148 | ± | 0.0019 | 10 | 39.1 | 0.0173 | ± | 0.0025 | 16 | 70.0 | 0.0220 | ± | 0.0016 |
| 5 | B | 1.57 | ± | 0.04 | 4 | 9.10 | 1.18 | ± | 0.02 | 7 | 14.0 | 1.33 | ± | 0.02 | 11 | 23.1 | 0.744 | ± | 0.172 |
| 6 | C | 42185 | ± | 719 | 9 | 11.9 | 45772 | ± | 2215 | 11 | 38.7 | 44928 | ± | 1702 | 20 | 50.6 | 41300 | ± | 8400 |
| 7 | N | 1473 | ± | 26 | 9 | 11.0 | 1439 | ± | 24 | 11 | 14.7 | 1454 | ± | 17 | 20 | 25.7 | 2500 | ± | 660 |
| 8 | O | | | | | | 395000 | ± | 2844 | 1 | 24.1 | 395000 | ± | 2844 | 1 | 24.1 | 453840 | ± | 20000 |
| 9 | F | | | | | | | | | | | | | | | | 92.0 | ± | 40.0 |
| 10 | Ne | 0.0141 | ± | 0.0021 | 19 | 11.6 | 0.000770 | ± | 0.000059 | 15 | 15.3 | 0.00650 | ± | 0.00090 | 34 | 26.9 | 0.000180 | | |
| 11 | Na | 5433 | ± | 79 | 9 | 36.2 | 6707 | ± | 76 | 11 | 39.7 | 6099 | ± | 54 | 20 | 75.9 | 5100 | ± | 500 |
| 12 | Mg | 109864 | ± | 1082 | 9 | 36.2 | 111505 | ± | 701 | 11 | 39.7 | 110722 | ± | 633 | 20 | 75.9 | 95170 | ± | 4000 |
| 13 | Al | 8987 | ± | 112 | 9 | 36.2 | 9263 | ± | 68 | 11 | 39.7 | 9131 | ± | 64 | 20 | 75.9 | 8370 | ± | 600 |
| 14 | Si | 131505 | ± | 1808 | 7 | 10.7 | 123880 | ± | 798 | 10 | 39.1 | 125520 | ± | 738 | 17 | 49.8 | 107740 | ± | 7200 |
| 15 | P | 1295 | ± | 37 | 8 | 34.6 | 1308 | ± | 17 | 10 | 39.1 | 1302 | ± | 19 | 18 | 73.6 | 978 | ± | 120 |
| 16 | S | 55059 | ± | 384 | 5 | 9.18 | 57481 | ± | 308 | 9 | 38.3 | 57013 | ± | 259 | 14 | 47.4 | 53600 | ± | 4400 |
| 17 | Cl | | | | | | 776 | ± | 21 | 1 | 24.1 | 776 | ± | 21 | 1 | 24.1 | 717 | ± | 270 |
| 18 | Ar | 0.0608 | ± | 0.0016 | 12 | 0.903 | 0.101 | ± | 0.003 | 6 | 0.408 | 0.0733 | ± | 0.0015 | 18 | 1.31 | 0.00133 | | |
| 19 | K | 522 | ± | 15 | 9 | 36.2 | 522 | ± | 9 | 11 | 39.7 | 522 | ± | 8 | 20 | 75.9 | 539 | ± | 48 |
| 20 | Ca | 15622 | ± | 288 | 9 | 36.2 | 12319 | ± | 116 | 10 | 39.1 | 13908 | ± | 151 | 19 | 75.3 | 8840 | ± | 700 |
| 21 | Sc | 6.58 | ± | 0.24 | 9 | 36.2 | 6.20 | ± | 0.06 | 11 | 39.7 | 6.38 | ± | 0.12 | 20 | 75.9 | 5.83 | ± | 0.40 |
| 22 | Ti | 485 | ± | 5 | 5 | 33.0 | 464 | ± | 8 | 8 | 38.2 | 474 | ± | 5 | 13 | 71.1 | 450 | ± | 30 |
| 23 | V | 57.7 | ± | 1.6 | 9 | 36.2 | 63.8 | ± | 2.7 | 11 | 39.7 | 60.9 | ± | 1.6 | 20 | 75.9 | 53.6 | ± | 4.0 |
| 24 | Cr | 2609 | ± | 35 | 6 | 34.6 | 2745 | ± | 21 | 9 | 38.8 | 2681 | ± | 20 | 15 | 73.4 | 2610 | ± | 200 |
| 25 | Mn | 2795 | ± | 30 | 9 | 36.2 | 2209 | ± | 17 | 11 | 39.7 | 2489 | ± | 17 | 20 | 75.9 | 1896 | ± | 160 |
| 26 | Fe | 198288 | ± | 1787 | 9 | 36.2 | 200972 | ± | 1301 | 11 | 39.7 | 199692 | ± | 1091 | 20 | 75.9 | 185620 | ± | 13000 |
| 27 | Co | 559 | ± | 9 | 9 | 36.2 | 602 | ± | 8 | 11 | 39.7 | 582 | ± | 5 | 20 | 75.9 | 508 | ± | 30 |
| 28 | Ni | 12014 | ± | 115 | 9 | 36.2 | 11745 | ± | 75 | 11 | 39.7 | 11873 | ± | 68 | 20 | 75.9 | 10950 | ± | 700 |
| 29 | Cu | 121 | ± | 3 | 8 | 34.6 | 131 | ± | 2 | 10 | 39.1 | 126 | ± | 2 | 18 | 73.6 | 130 | ± | 20 |
| 30 | Zn | 347 | ± | 7 | 9 | 36.2 | 344 | ± | 4 | 11 | 39.7 | 346 | ± | 4 | 20 | 75.9 | 311 | ± | 20 |
| 31 | Ga | 10.0 | ± | 0.4 | 9 | 36.2 | 10.0 | ± | 0.6 | 11 | 39.7 | 10.0 | ± | 0.4 | 20 | 75.9 | 9.45 | ± | 0.70 |
| 32 | Ge | 34.8 | ± | 0.6 | 3 | 7.87 | 32.5 | ± | 1.4 | 4 | 31.8 | 33.0 | ± | 1.1 | 7 | 39.7 | 33.4 | ± | 3.0 |
| 33 | As | 1.81 | ± | 0.04 | 5 | 33.4 | 1.97 | ± | 0.05 | 5 | 32.5 | 1.89 | ± | 0.03 | 10 | 65.9 | 1.77 | ± | 0.16 |
| 34 | Se | 22.5 | ± | 0.8 | 5 | 33.4 | 24.8 | ± | 0.7 | 5 | 32.5 | 23.6 | ± | 0.5 | 10 | 65.9 | 20.4 | ± | 1.6 |
| 35 | Br | | | | | | | | | | | | | | | | 3.77 | ± | 1.80 |
| 36 | Kr | 0.000177 | ± | 0.000002 | 12 | 0.903 | 0.000246 | ± | 0.000005 | 5 | 0.240 | 0.000191 | ± | 0.000002 | 17 | 1.14 | 0.0000522 | | |
| 37 | Rb | 2.33 | ± | 0.06 | 8 | 34.6 | 2.41 | ± | 0.07 | 10 | 39.1 | 2.37 | ± | 0.05 | 18 | 73.6 | 2.22 | ± | 0.18 |
| 38 | Sr | 12.7 | ± | 0.2 | 8 | 34.6 | 10.9 | ± | 0.3 | 10 | 39.1 | 11.8 | ± | 0.2 | 18 | 73.6 | 7.79 | ± | 0.50 |
| 39 | Y | 2.09 | ± | 0.03 | 8 | 34.6 | 1.64 | ± | 0.03 | 10 | 39.1 | 1.85 | ± | 0.02 | 18 | 73.6 | 1.50 | ± | 0.10 |
| 40 | Zr | 3.90 | ± | 0.07 | 5 | 33.0 | 4.52 | ± | 0.08 | 8 | 38.2 | 4.23 | ± | 0.05 | 13 | 71.1 | 3.79 | ± | 0.28 |
| 41 | Nb | 0.386 | ± | 0.006 | 5 | 33.0 | 0.355 | ± | 0.004 | 8 | 38.2 | 0.369 | ± | 0.004 | 13 | 71.1 | 0.279 | ± | 0.015 |
| 42 | Mo | 0.987 | ± | 0.014 | 5 | 33.0 | 1.05 | ± | 0.01 | 8 | 38.2 | 1.02 | ± | 0.01 | 13 | 71.1 | 0.976 | ± | 0.050 |
| 44 | Ru | 0.551 | ± | 0.030 | 3 | 7.87 | 0.643 | ± | 0.018 | 4 | 31.8 | 0.625 | ± | 0.016 | 7 | 39.7 | 0.666 | ± | 0.040 |
| 45 | Rh | | | | | | 0.150 | ± | 0.010 | 1 | 24.1 | 0.150 | ± | 0.010 | 1 | 24.1 | 0.133 | ± | 0.008 |
| 46 | Pd | 0.503 | ± | 0.010 | 3 | 7.87 | 0.804 | ± | 0.009 | 4 | 31.8 | 0.744 | ± | 0.007 | 7 | 39.7 | 0.558 | ± | 0.030 |
| 47 | Ag | 0.195 | ± | 0.009 | 1 | 23.9 | 0.195 | ± | 0.006 | 1 | 24.1 | 0.195 | ± | 0.006 | 2 | 48.0 | 0.204 | ± | 0.008 |
| 48 | Cd | 0.706 | ± | 0.023 | 8 | 34.6 | 0.744 | ± | 0.015 | 10 | 39.1 | 0.726 | ± | 0.014 | 18 | 73.6 | 0.679 | ± | 0.024 |
| 49 | In | 0.103 | ± | 0.004 | 3 | 29.8 | 0.0906 | ± | 0.0026 | 4 | 30.0 | 0.0970 | ± | 0.0023 | 7 | 59.8 | 0.0786 | ± | 0.0040 |
| 50 | Sn | 1.31 | ± | 0.03 | 4 | 9.10 | 1.98 | ± | 0.13 | 8 | 38.2 | 1.85 | ± | 0.11 | 12 | 47.3 | 1.63 | ± | 0.16 |
| 51 | Sb | 0.156 | ± | 0.007 | 5 | 10.7 | 0.174 | ± | 0.006 | 8 | 14.7 | 0.166 | ± | 0.005 | 13 | 25.4 | 0.169 | ± | 0.018 |
| 52 | Te | 2.19 | ± | 0.05 | 3 | 7.87 | 2.39 | ± | 0.15 | 4 | 31.8 | 2.35 | ± | 0.12 | 7 | 39.7 | 2.31 | ± | 0.18 |
| 53 | I | | | | | | | | | | | | | | | | 0.770 | ± | 0.620 |
| 54 | Xe | 0.000560 | ± | 0.000004 | 14 | 0.907 | 0.000616 | ± | 0.000009 | 6 | 0.251 | 0.000572 | ± | 0.000004 | 20 | 1.16 | 0.000174 | | |
| 55 | Cs | 0.212 | ± | 0.003 | 8 | 34.6 | 0.201 | ± | 0.004 | 10 | 39.1 | 0.206 | ± | 0.002 | 18 | 73.6 | 0.188 | ± | 0.012 |
| 56 | Ba | 3.38 | ± | 0.06 | 8 | 34.6 | 2.90 | ± | 0.04 | 10 | 39.1 | 3.12 | ± | 0.04 | 18 | 73.6 | 2.39 | ± | 0.16 |
| 57 | La | 0.303 | ± | 0.004 | 8 | 34.6 | 0.275 | ± | 0.003 | 10 | 39.1 | 0.288 | ± | 0.002 | 18 | 73.6 | 0.244 | ± | 0.016 |
| 58 | Ce | 0.785 | ± | 0.010 | 8 | 34.6 | 0.664 | ± | 0.014 | 10 | 39.1 | 0.721 | ± | 0.009 | 18 | 73.6 | 0.627 | ± | 0.052 |
| 59 | Pr | 0.124 | ± | 0.002 | 8 | 34.6 | 0.109 | ± | 0.002 | 10 | 39.1 | 0.116 | ± | 0.001 | 18 | 73.6 | 0.0951 | ± | 0.0066 |
| 60 | Nd | 0.666 | ± | 0.013 | 8 | 34.6 | 0.540 | ± | 0.020 | 10 | 39.1 | 0.599 | ± | 0.012 | 18 | 73.6 | 0.472 | ± | 0.036 |
| 62 | Sm | 0.208 | ± | 0.004 | 8 | 34.6 | 0.182 | ± | 0.005 | 10 | 39.1 | 0.194 | ± | 0.003 | 18 | 73.6 | 0.153 | ± | 0.012 |
| 63 | Eu | 0.0818 | ± | 0.0011 | 8 | 34.6 | 0.0684 | ± | 0.0012 | 10 | 39.1 | 0.0747 | ± | 0.0008 | 18 | 73.6 | 0.0577 | ± | 0.0050 |
| 64 | Gd | 0.303 | ± | 0.005 | 8 | 34.6 | 0.246 | ± | 0.004 | 10 | 39.1 | 0.273 | ± | 0.003 | 18 | 73.6 | 0.208 | ± | 0.018 |
| 65 | Tb | 0.0541 | ± | 0.0007 | 8 | 34.6 | 0.0451 | ± | 0.0006 | 10 | 39.1 | 0.0493 | ± | 0.0005 | 18 | 73.6 | 0.0380 | ± | 0.0030 |
| 66 | Dy | 0.375 | ± | 0.004 | 8 | 34.6 | 0.313 | ± | 0.004 | 10 | 39.1 | 0.342 | ± | 0.003 | 18 | 73.6 | 0.252 | ± | 0.020 |
| 67 | Ho | 0.0823 | ± | 0.0011 | 8 | 34.6 | 0.0677 | ± | 0.0006 | 10 | 39.1 | 0.0745 | ± | 0.0006 | 18 | 73.6 | 0.0563 | ± | 0.0044 |
| 68 | Er | 0.247 | ± | 0.004 | 8 | 34.6 | 0.206 | ± | 0.003 | 10 | 39.1 | 0.225 | ± | 0.002 | 18 | 73.6 | 0.164 | ± | 0.012 |
| 69 | Tm | 0.0381 | ± | 0.0006 | 8 | 34.6 | 0.0316 | ± | 0.0006 | 10 | 39.1 | 0.0346 | ± | 0.0004 | 18 | 73.6 | 0.0259 | ± | 0.0024 |
| 70 | Yb | 0.247 | ± | 0.003 | 8 | 34.6 | 0.205 | ± | 0.003 | 10 | 39.1 | 0.225 | ± | 0.002 | 18 | 73.6 | 0.167 | ± | 0.014 |
| 71 | Lu | 0.0368 | ± | 0.0009 | 8 | 34.6 | 0.0313 | ± | 0.0006 | 10 | 39.1 | 0.0339 | ± | 0.0005 | 18 | 73.6 | 0.0249 | ± | 0.0020 |
| 72 | Hf | 0.120 | ± | 0.003 | 5 | 33.0 | 0.124 | ± | 0.007 | 8 | 38.2 | 0.122 | ± | 0.004 | 13 | 71.1 | 0.106 | ± | 0.008 |
| 73 | Ta | 0.0460 | ± | 0.0009 | 5 | 33.0 | 0.474 | ± | 0.007 | 8 | 38.2 | 0.275 | ± | 0.004 | 13 | 71.1 | 0.0148 | ± | 0.0014 |
| 74 | W | 0.106 | ± | 0.004 | 5 | 33.0 | 0.142 | ± | 0.003 | 8 | 38.2 | 0.125 | ± | 0.002 | 13 | 71.1 | 0.102 | ± | 0.014 |
| 75 | Re | 0.0364 | ± | 0.0019 | 3 | 7.87 | 0.0463 | ± | 0.0024 | 3 | 7.69 | 0.0413 | ± | 0.0015 | 6 | 15.6 | 0.0369 | ± | 0.0028 |
| 76 | Os | 0.461 | ± | 0.019 | 4 | 9.50 | 0.494 | ± | 0.011 | 4 | 8.34 | 0.476 | ± | 0.011 | 8 | 17.8 | 0.475 | ± | 0.020 |
| 77 | Ir | 0.449 | ± | 0.008 | 4 | 9.50 | 0.542 | ± | 0.011 | 4 | 8.34 | 0.492 | ± | 0.007 | 8 | 17.8 | 0.474 | ± | 0.020 |
| 78 | Pt | 0.767 | ± | 0.021 | 3 | 7.87 | 0.891 | ± | 0.025 | 3 | 7.69 | 0.828 | ± | 0.017 | 6 | 15.6 | 0.931 | ± | 0.072 |
| 79 | Au | 0.193 | ± | 0.002 | 1 | 1.63 | 0.0737 | ± | 0.0016 | 1 | 0.648 | 0.159 | ± | 0.002 | 2 | 2.28 | 0.147 | ± | 0.024 |
| 80 | Hg | | | | | | | | | | | | | | | | 0.288 | ± | 0.140 |
| 81 | Tl | 0.117 | ± | 0.004 | 8 | 34.6 | 0.121 | ± | 0.005 | 10 | 39.1 | 0.119 | ± | 0.003 | 18 | 73.6 | 0.141 | ± | 0.014 |
| 82 | Pb | 2.75 | ± | 0.08 | 8 | 34.6 | 2.93 | ± | 0.07 | 10 | 39.1 | 2.84 | ± | 0.05 | 18 | 73.6 | 2.64 | ± | 0.16 |
| 83 | Bi | 0.109 | ± | 0.003 | 8 | 34.6 | 0.112 | ± | 0.004 | 10 | 39.1 | 0.110 | ± | 0.002 | 18 | 73.6 | 0.113 | ± | 0.016 |
| 90 | Th | 0.0332 | ± | 0.0011 | 8 | 34.6 | 0.0335 | ± | 0.0009 | 10 | 39.1 | 0.0334 | ± | 0.0007 | 18 | 73.6 | 0.0298 | ± | 0.0030 |
| 92 | U | 0.00924 | ± | 0.00034 | 8 | 34.6 | 0.00920 | ± | 0.00048 | 10 | 39.1 | 0.00922 | ± | 0.00030 | 18 | 73.6 | 0.00816 | ± | 0.00106 |

(Bizzarro *et al.*, 2023), K (Hu *et al.*, 2024), Ca (Moynier *et al.*, 2022), Cr and Ti (Yokoyama *et al.*, 2023b), Fe (Hopp *et al.*, 2022), Cu and Zn (Paquet *et al.*, 2023), Ni (Spitzer *et al.*, 2024), Zr (Schönbächler *et al.*, 2024), Mo (Nakanishi *et al.*, 2023), and Sm-Nd (Torrano *et al.*, 2024). Similar to the results of elemental abundances, most of the isotopic compositions in bulk Ryugu samples were generally consistent with those of CI chondrites (see references above), suggesting that the source materials of Ryugu and CIs share a common genetic heritage. In particular, Ryugu and CIs have indistinguishable $^{54}Fe/^{56}Fe$ ratios, which are different from the other carbonaceous chondrite (CC) and non-carbonaceous (NC) chondrite groups (Hopp *et al.*, 2022). A hypothetical explanation by Hopp *et al*. (2022) is that the parent bodies of Ryugu and CIs were deflected into the main belt from the outer Solar System by excitation of Uranus and Neptune, while other CC bodies formed in more inner regions of the Solar System, near Jupiter and Saturn. In contrast, $^{54}Cr/^{52}Cr$ ratios in Ryugu samples (and CI chondrites) with masses <25 mg showed a variation exceeding the documented dispersion of literature values for CIs, whereas the calculated $^{54}Cr/^{52}Cr$ ratio of a total of ~90 mg of the bulk Ryugu sample is consistent with the CI value (Yokoyama *et al.*, 2023b). This observation suggests the presence of sub-mm to mm-scale $^{54}Cr/^{52}Cr$ variability in the bulk Ryugu samples (and possibly in CIs), which could have primarily been caused by fluid-driven parent body aqueous alteration. Yokoyama et al. (2023b) argued that pervasive aqueous alteration in the Ryugu parent body released Cr with relatively low $^{54}Cr/^{52}Cr$ ratios from chemically labile phases (e.g., olivine and amorphous silicates) whereas $^{54}Cr$-rich presolar grains were largely unaffected, resulting in the incorporation of $^{54}Cr$-poor Cr dissolved in the aqueous fluid into secondary minerals such as dolomite, magnetite, and pyrrhotite. On the other hand, Ryugu showed greater s-process depletion of Mo isotopes compared to any known bulk CCs, including CIs (Nakanishi *et al.*, 2023). The different Mo isotopic compositions of Ryugu and CIs could be caused by biased sampling of Ryugu materials. Ryugu is enriched in aqueously formed secondary minerals with s-process-poor Mo isotopes, resulting from the physicochemical separation of s-process-rich presolar grains and a complementary s-process-poor aqueous fluid in the Ryugu parent body. However, incomplete digestion of s-process-rich presolar SiC during dissolution of Ryugu samples in the laboratory cannot be ruled out. The small-scale isotopic heterogeneity of Ryugu, as suggested by the two case studies on Cr and Mo isotopes, is most likely related to the variability of elemental abundances in Ryugu samples due to the heterogeneous distribution of minor secondary minerals, as discussed below.

## 3. Evaluation of Ryugu sample heterogeneity

### 3.1. Variation of elemental abundances in Ryugu samples

**Figure 2** shows the CI-normalized abundances of 18 selected elements in the Ryugu samples (Ito *et al.*, 2022; Nakamura *et al.*, 2022; Yokoyama *et al.*, 2023a). The diamond data points are data measured by XRF in Yokoyama *et al.* (2023a) using C0108 (24 mg), and the circle symbols are those obtained by ICP-MS measurements in the same study using A0106 and A0107 combined (24 mg) and C0108 (pre-measured by XRF). The ICP-MS results obtained from the two samples are generally in good agreement, with the maximum relative percentage difference (RPD = $|X_{A0106+A0107}-X_{C0108}|$ / $(X_{A0106+A0107}/2+X_{C0108}/2) \times 100$) of 22% for P. The box-plots in Fig. 2a show the variation of analytical data for 0.33–3.3 mg of Ryugu samples (n = 18: Ito *et al.*, 2022; Nakamura *et al.*, 2022), which show large dispersions for some elements, including P (0.2–2.1 × CI), Ca (0.6–2.8 × CI), Mn (0.5–2.3 × CI), Sr (0.6–2.8 × CI), La (0.4–1.5 × CI), Gd (0.3–2.2 × CI), and Yb (0.1–2.4 × CI) compared with the other elements that show dispersions mostly in the range of 0.8–1.4 × CI . Specifically, samples from Chamber A (*n* = 8) show larger dispersions in these elements than samples from Chamber C (*n* = 10) (**Figs. 2b, 2c**).

The observed variations in some elements for the measurements of relatively small samples (< 3.3 mg) most likely stem from the presence of aqueously formed secondary minerals in Ryugu (e.g., carbonate and phosphate), in which some specific elements, including P, Mn, Sr, and rare earth elements (REEs), are strongly partitioned when these minerals precipitate. The modal abundances of carbonates (0.43–6.93%) and phosphates (0.16–1.88%), as well as magnetite (0.47–6.47%) and Fe-sulfide (1.11–5.24%), are relatively low and vary from fragment to fragment in Ryugu (Nakamura *et al.*, 2022). Therefore, the elemental abundances obtained for small-sized samples are controlled by the amount of relevant minerals involved (i.e., nugget effect). Similar variability can be observed for the abundances of P and Ca in fragments of CI chondrites, resulting from the different mineralogical compositions of the measured fragments (Morlok *et al.*, 2006). Additionally, measurements of ~40 mg Ivuna samples by King *et al*. (2020) showed inconsistent REE abundances compared to those of a 0.71 g Ivuna sample (Barrat *et al.*, 2012). Even in the largest 24 mg Ryugu sample examined to date, the influence of carbonates and phosphates can be seen for P, Ca, and Mn (**Fig. 2**), as the two bulk samples (TD1 and TD2) showed ~20% of RPD in the abundances of these elements. Below, we discuss the influence of nugget minerals on the elemental abundances of Ryugu more in detail.

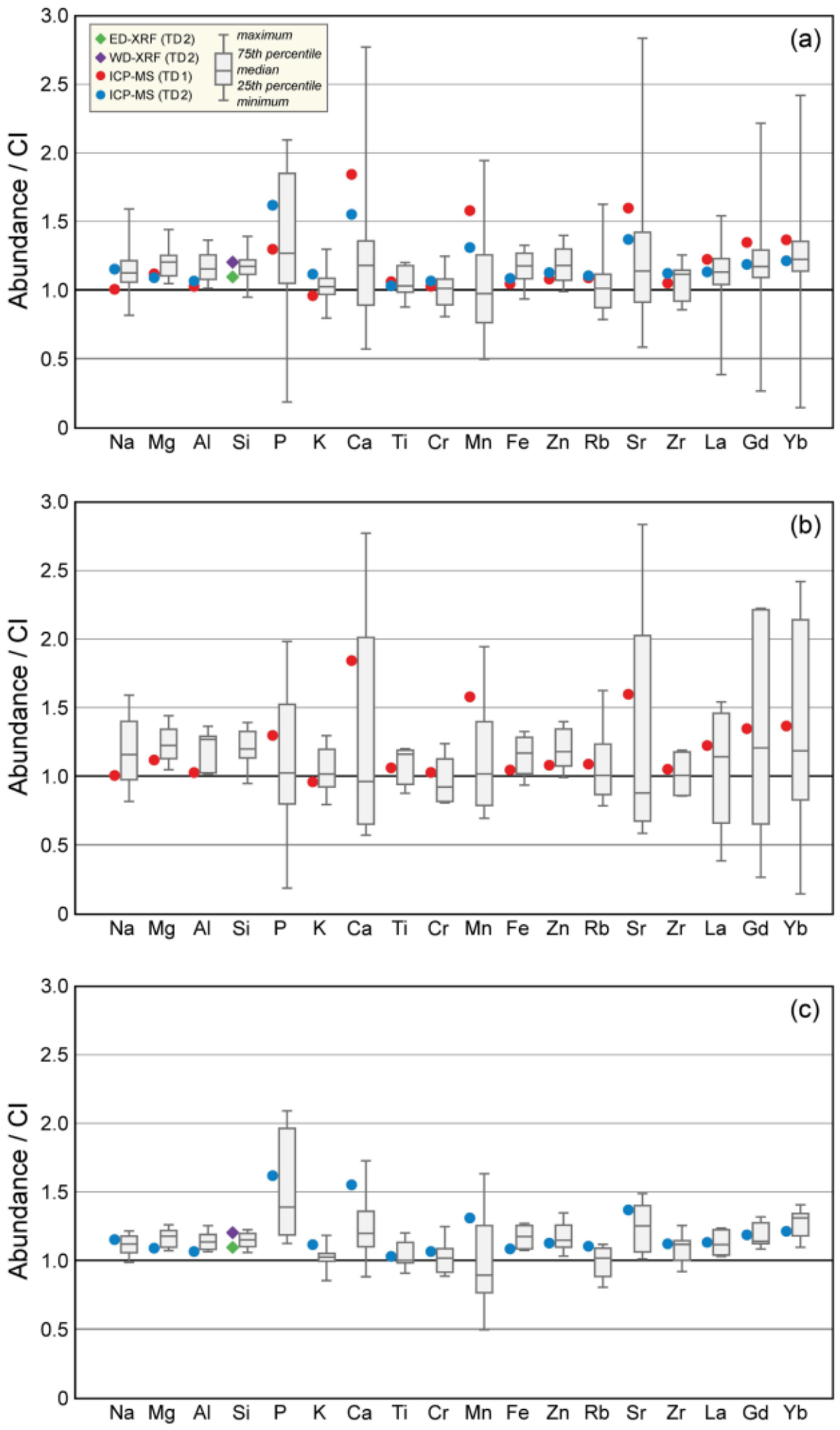

**Fig. 2** CI-normalized elemental abundances in bulk Ryugu samples from (a) Chambers A ("TD1") and C ("TD2") combined, (b) Chamber A, and (c) Chamber C. The diamonds (Energy Dispersive (ED)- and Wavelength Dispersive (WD)-XRF) and circles (ICP-MS) represent data obtained from the measurements of ~25 mg powdered Ryugu samples (Yokoyama et al., 2023a). Boxplots were produced from the ICP-MS data of 16 individual Ryugu grains with digested masses of 0.2–3 mg, except for Si, which was measured by electron probe (Nakamura et al., 2022).

## 3.2. Nugget effect of secondary minerals

There are several telltale characteristics of the nugget effect of minor secondary minerals. In particular, "bulk" analyses of relatively small samples should define mixing lines between the "true bulk" and the mineral composition in elemental ratio plots. Here we evaluate the extent of variation for elemental abundance ratios in Ryugu samples associated with the presence of four minor mineral phases mentioned in the previous section: dolomite; $CaMg(CO_3)_2$, apatite; $Ca_5(PO_4)_3(OH,F,Cl)$, magnetite; $Fe_3O_4$, and pyrrhotite; $Fe_{1-x}S$. Other carbonates (e.g., breunnerite; $(Mg,Fe)CO_3$, calcite; $CaCO_3$), phosphates (e.g., Na-Mg phosphate), oxides (e.g., ilmenite; $FeTiO_3$, chromite; $(Fe,Mg)Cr_2O_4$), and sulfides (e.g., pentlandite; $(Fe,Ni)_9S_8$) found in Ryugu samples are less abundant compared to the abovementioned minerals (Nakamura *et al.*, 2023), and are thus not considered here. Key elements used in the following discussion include P, S, Ca, Ti, Cr, Mn, Fe, and Sr, of which the abundances in dolomite, apatite, magnetite, and pyrrhotite from Ryugu were taken from Bazi *et al*. (2022), Nakamura *et al*. (2022), and Nakamura *et al*. (2023) (**Table 2**). As shown in **Table 2**, these elements are selectively distributed in individual minerals: Ca-Mn-Sr for dolomite, Ca-P-Sr for apatite, Fe-Ti for magnetite, and Fe-S-Ti for pyrrhotite. In contrast, the abundance of Cr is found to be less variable among these minerals (1500–3000 μg/g), which is similar to the bulk Ryugu value (2681 μg/g, Table 1). Therefore, we take Cr as the common denominator and evaluate the variations of the CI-normalized ratios of $(P/Cr)_N$, $(S/Cr)_N$, $(Ca/Cr)_N$, $(Ti/Cr)_N$, $(Mn/Cr)_N$, $(Fe/Cr)_N$, and $(Sr/Cr)_N$.

**Figures 3a-3c** show the $(Mn/Cr)_N$, $(Sr/Cr)_N$, and $(P/Cr)_N$ ratios of individual Ryugu samples plotted against the $(Ca/Cr)_N$ ratio. Also shown in these figures are mixing lines representing the addition/removal of dolomite and apatite to/from the bulk CI chondrite, while those for magnetite and pyrrhotite are not shown since these elements are less sensitive to variations in Ca-Mn-P-Sr abundances in Ryugu samples. In **Figs. 3a and 3b**, the $(Mn/Cr)_N$ and $(Sr/Cr)_N$ ratios of Ryugu samples are positively correlated with the $(Ca/Cr)_N$ ratio, generally following the CI-dolomite mixing line from CI–2.5% to CI+10%, with slight shift toward the direction of CI-apatite mixing line. A similar positive correlation is observed between $(P/Cr)_N$ and $(Ca/Cr)_N$ ratios, with data points scattered between the CI-dolomite and CI-apatite mixing lines (**Figs. 3c**). These observations indicate that dolomite is the major contributor controlling the abundances of Ca-Mn-Sr in individual Ryugu samples. The impact of apatite on the $(Mn/Cr)_N$ ratio is not as pronounced as that of dolomite for a given modal abundance, due to the relatively low Mn abundance in apatite (**Table 2**). Nevertheless, a non-negligible influence of apatite (up to CI+1%) is confirmed for varying the abundances of P, Sr, and most likely REEs. Interestingly, TD1 samples show relatively higher REE abundances than TD2 samples, whereas both bulk samples have nearly the same P abundances (**Fig. 1**). As shown in **Fig. 3c**, some TD1 samples show greater contribution

**Table 2** Elemental abundances of secondary minerals

| Element | Dolomite | | Apatite | | Magnetite | | Pyrrhotite | |
|---|---|---|---|---|---|---|---|---|
| | µg/g | References | µg/g | References | µg/g | References | µg/g | References |
| P | | | 173581 | B22, N23 | | | | |
| S | | | 7900 | B22 | | | 413300 | B22 |
| Ca | 187274 | B22, N23 | 367876 | B22, N23 | | | | |
| Ti | 350 | B22 | | | 3633 | B22 | 4400 | B22 |
| Cr | 2047 | B22 | 1500 | B22 | 2367 | B22 | 3000 | B22 |
| Mn | 45232 | B22, N23 | 5280 | B22, N23 | 508 | B22 | 547 | B22 |
| Fe | 17779 | B22, N23 | 14405 | B22, N23 | 723600 | B22 | 539800 | B22 |
| Sr | 98 | B22, N22 | 522 | B22, N22 | | | | |

*References: B22 (Bazi et al., 2022); M84 (Macdougall et al., 1984), N22 (Nakamura et al., 2022).*

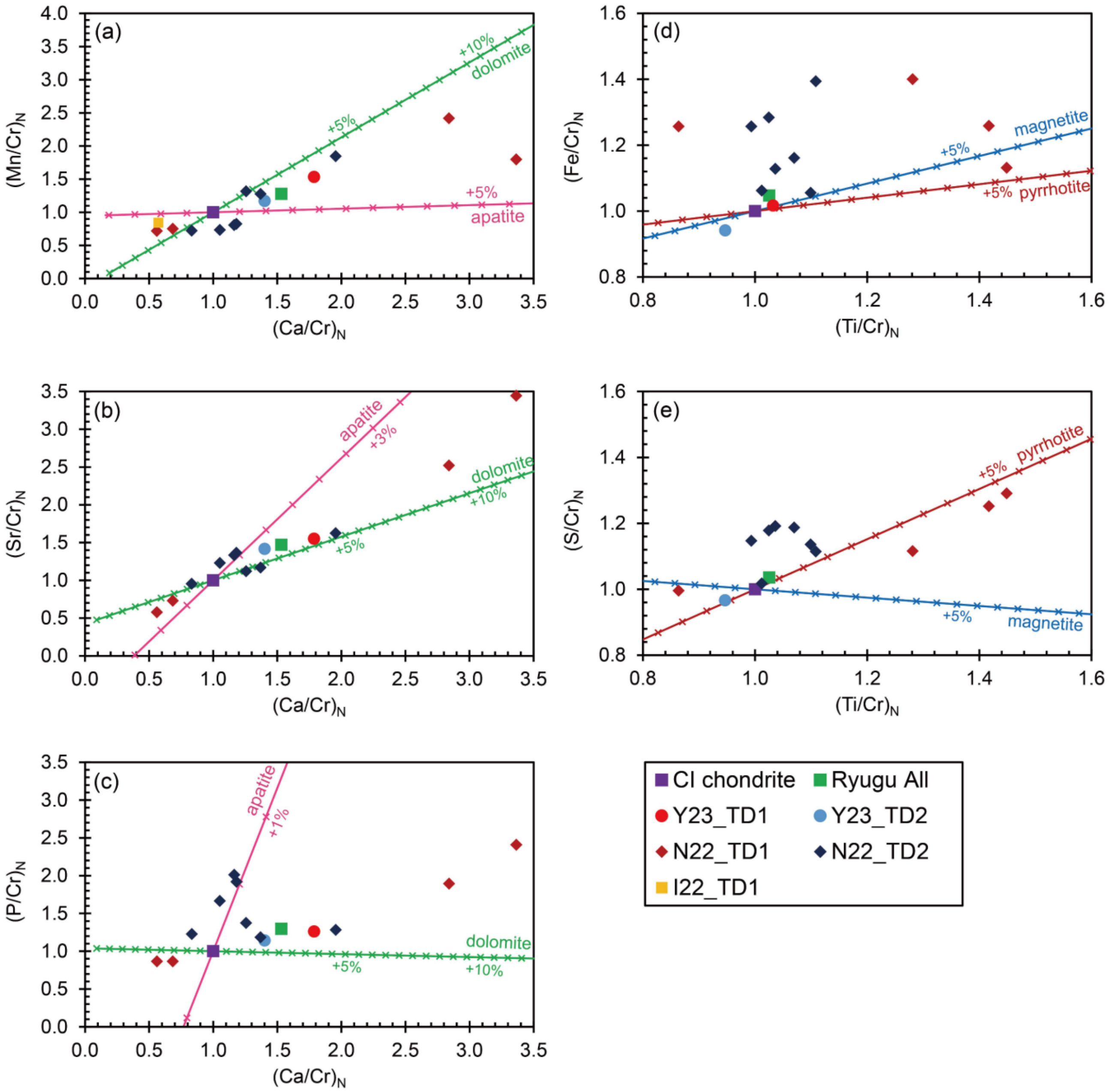

**Fig. 3** Correlations between CI chondrite-normalized ratios of (a) Mn/Cr–Ca/Cr, (b) Fe/Cr–Ca/Cr, (c) P/Cr–Ca/Cr, (d) Fe/Cr–Ti/Cr, and (e) S/Cr–Ti/Cr for bulk Ryugu samples obtained by Yokoyama et al. (2023a) ("Y23"), Nakamura et al. (2022) ("N22"), and so on I22 (Ito et al., 2022). Purple squares are CI chondrite composition (Lodders, 2021) and green squares are average Ryugu composition (). The lines represent the addition or removal of dolomite (green), apatite (pink), magnetite (blue), and pyrrhotite (brown) relative to the bulk CI chondrite.

of dolomite than TD2 samples with similar contribution of apatite. Thus, the observed high REE abundances in TD1 samples may be due to either the influence of dolomite or other mineral phases that are not discussed here. Notably, the mass-weighted mean composition of Ryugu calculated from all measured data (green square) is distinct from the CI chondrite (purple square), indicating an approximately 1–2% and 0.5% enrichment of dolomite and apatite, respectively, relative to the CI chondrite.

**Figures 3d-3e** show the $(Fe/Cr)_N$ and $(S/Cr)_N$ ratios of Ryugu samples plotted against the $(Ti/Cr)_N$ ratio, together with the CI-magnetite and CI-pyrrhotite mixing lines. The mixing lines for CI-dolomite and CI-apatite are not presented in these figures as Fe, S, and Ti would not be affected by the distribution of these minerals in the Ryugu samples. The Ryugu data points deviate from the two mixing lines in **Fig. 3d**, which are difficult to explain by the systematic enrichment or depletion of magnetite and pyrrhotite. Morlok *et al*. (2006) measured the elemental abundances of fragments in the size range from 40 to several hundred µm in CI chondrites, and argued that the compositional variation of major elements (e.g., Fe, Mg, Si) in the CI chondrite fragments reflects their different lithologies due to the heterogeneities in the starting material before the onset of aqueous alteration. Therefore, the Fe abundance in individual Ryugu samples is presumably not sensitive to the amount of minor minerals, but dominantly controlled by the composition of phyllosilicates (e.g., serpentine, saponite) consisting of the Ryugu matrix (Nakamura *et al.*, 2023). In contrast, the positive correlation between $(S/Cr)_N$ and $(Ti/Cr)_N$ for individual Ryugu samples suggests the incorporation of pyrrhotite (**Fig 3e**). Again, the mass-weighted mean composition of Ryugu is distinct from the CI chondrite and ~0.5% enrichments of pyrrhotite is expected.

In summary, carbonates appear to be the most important mineral whose modal abundance largely controls the abundances of Ca-Mn-Sr in individual Ryugu samples. The nugget effect of phosphate and pyrrhotite is also confirmed, affecting the abundances of P-Sr and S-Ti of some Ryugu samples, respectively. The abundance of Fe is less sensitive to the nugget effect of minor minerals including magnetite, of which the variation in individual Ryugu samples is due to the difference in lithology of the fragments, reflecting the heterogeneities of original material prior to the aqueous alteration.

### 3.3. Statistical analysis of the nugget effect

Another telltale signature of a nugget effect is that the dispersion in elemental ratios should decrease as the inverse of the square root of the mass of the homogenized sample (Dauphas and Pourmand, 2015). Such mass-dependent dispersion in elemental ratios is consistent with the sampling problem associated with the nugget effect. Considering two elements with mass fractions $C_1$ and $C_2$ in a sample of mass = $m$, the dispersion of the elemental ratio $R = C_2/C_1$ can be calculated by the following equation (Dauphas and Pourmand, 2015);

$$\sigma_R \approx \frac{r(\rho_{\text{nugget}}/\rho_{\text{matrix}})}{[1 + rf(\rho_{\text{nugget}}/\rho_{\text{matrix}})]^2} \sqrt{\frac{f\rho_{\text{matrix}}\pi d^3}{6m}} \times |R_{\text{nugget}} - R_{\text{matrix}}| \quad (3)$$

where $r$ is $C_{1_nugget}/C_{1_matrix}$, $\rho$ is the density, $f$ is the volume fraction of nuggets, and $d$ is the particle diameter of the nugget. **Figure 4** shows the Mn/Cr ratios of individual Ryugu samples measured in previous studies, plotted against the digested mass. The red curves in this figure show the estimated dispersion (±2σ) of the Mn/Cr ratio calculated from equation (3) as a function of the sample mass used.

$$R(m) = \bar{R} \pm 2\sigma_R(m) \quad (4)$$

where $R(m)$ is the estimated maximum and minimum Mn/Cr ratio of the bulk Ryugu of mass $m$, and $\bar{R}$ is the mass-weighted mean Mn/Cr ratio of bulk Ryugu (TD1 + TD2) listed in **Table 1**. It is assumed here that carbonate is the only nugget phase, in which case the matrix is represented by subtracting carbonates from the bulk CI chondrite. The elemental abundances of Mn and Cr in Ryugu carbonates are given in **Table 2**, with the volume fraction of $f$ = 0.025 (Nakamura *et al.*, 2022). Then, the elemental abundances of Mn and Cr in the matrix (i.e., the non-carbonate phase) are calculated by subtracting 2.5% carbonates from the bulk CI chondrite (**Table 1**). The other parameters used in the calculation are: $\rho_{\text{nugget}}$ = 2.84 g/cm$^3$, $\rho_{\text{matrix}}$ = 2.587 g/cm$^3$ (grain density: Nakamura *et al.*, 2022), and $d$ = 0.10 mm.

As shown in **Fig. 4**, the dispersions of the CI-normalized Mn/Cr ratio for the individual Ryugu data, represented as $\log_{10}(Mn/Cr)_N$, are significantly larger than the dispersion estimated from equations (3) and (4) (thin red curves). Note that the thin red curves represent the dispersion of the Mn/Cr ratio that would be expected when a given mass *m* of powder (*x* axis of **Fig. 4**) is taken from a homogeneous powder (*d* = 0.10 mm) prepared from a sufficiently large amount of Ryugu sample with a given mineral modal abundance (carbonate = 2.5%: i.e., *f* = 0.025). Therefore, the observed inconsistency is caused by an inappropriate assumption of the parameters used in the calculation and/or the presence of additional nugget phases other than carbonates (e.g., phosphates). In particular, increasing the *d* and *f* values of carbonates in equation (3) monotonically increases the $\sigma_R$ value; changing the *d* value from 0.10 to 0.16 mm makes the $\sigma_R$ value twice as high, although such a large carbonate grain rarely exists in Ryugu (Nakamura *et al.*, 2022). Thus, the observed inconsistency suggests a non-uniform distribution of various nugget minerals of different sizes (*d*) at the mm-scale of the Ryugu particles examined in the elemental abundance measurements by Nakamura *et al*. (2022) and Ito *et al*. (2022).

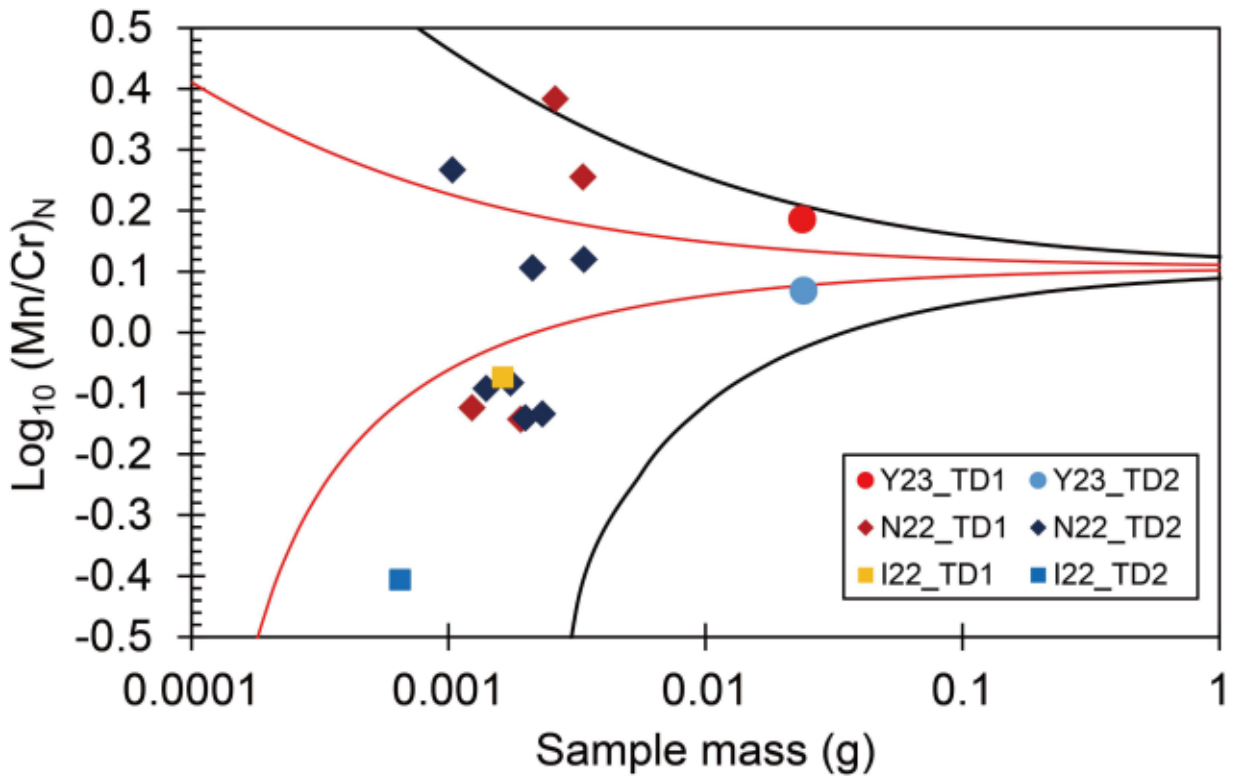

**Fig. 4** Elemental ratio of Mn/Cr plotted against sample mass. Symbols are the same as in **Fig. 3**. In the presence of a nugget effect, the dispersion of elemental ratios is expected to decrease as the inverse of the square root of the mass of the homogenized/digested sample (red thin curves) (Dauphas and Pourmand, 2015). However, this theoretical prediction underestimates the actual dispersion of the Ryugu data, due to non-uniform distribution of nugget minerals of different sizes within the mm-scale of the Ryugu particles. The bold black curves are the estimated data dispersion calculated by using the dispersion of data obtained from smaller Ryugu particles (< 3.3 mg, see main text for references).

To overcome this difficulty, we instead calculated the dispersion for the smaller particles and calculated the predicted dispersion for larger masses. Having determined that unrepresentative sampling of carbonates is likely responsible for the dispersion in certain elemental ratios, we can predict the expected dispersion when large sample masses are digested as follows;

$$\sigma_R(M) = \sigma_R(\bar{m})\sqrt{\bar{m}/M} \qquad (5)$$

where *M* is the expected sample mass to be digested, $\bar{m}$ is the average mass of the digested smaller particles, and $\sigma_R(\bar{m})$ is the standard deviation of *R* for the measurement of the smaller particles. Here, $\bar{m}$ and $\sigma_R(\bar{m})$ are determined using the data obtained in the measurements of smaller particles (< 3.3 mg) by Nakamura *et al.* (2022) and Ito *et al*. (2022). The following equation then gives the possible range of *R* in a Ryugu analysis digesting a homogenized sample of mass *M*;

$$R(M) = \bar{R} \pm 2\sigma_R(M) \qquad (6)$$

where $R(M)$ is the estimated maximum and minimum Mn/Cr ratio of the bulk Ryugu of mass *M*.

The bold black curves in **Fig. 4**, calculated from equations (5) and (6) (Dauphas and Pourmand, 2015), show the dispersion (2σ) of the Mn/Cr ratio for a Ryugu measurement assuming a sample mass *M*. It should be noted that the calculated dispersion corresponds to the maximum range of difference in the Mn/Cr ratio of the measured sample (mass = *M*) from those of the “true” bulk Ryugu sample, and thus represents the “analytical error” of Mn/Cr ratio. When the homogenized Ryugu sample weighs 0.1 g, the expected errors for the Mn/Cr ratio is ±13%. This level of analytical error is much larger than the measurement uncertainties of elemental ratios determined by a common ICP-MS instrument, which are better than ±3%. Homogenization of more samples reduces the expected error of elemental ratios. To achieve within ±5% error for Mn/Cr, the required amount of Ryugu sample to be homogenized is 0.7 g. To further improve the analytical error better than ±3%, >2 g of homogenized Ryugu sample is required.

A similar attempt was made for CI-normalized elemental abundance, $(X)_N$. **Figure 5** compares the extent of dispersion in the Ryugu data for some elements discussed in the previous section (P, Ca, Ti, Mn, Fe, and Sr), plotting their $\log_{10}(X)_N$ values. The extent of dispersion increases in the order of Fe < Ti < Mn < Ca ≈ Sr ≈ P, for which the expected amounts of homogenized Ryugu sample to obtain the $(X)_N$ values within ±5% error are 26, 41, 197, 335, 347, and 387 mg, respectively. This outcome is consistent with the nugget effect observed for carbonate (Mn, Ca, and Sr) and phosphate (P). However, these expected masses were calculated based on the dispersion of data for *all* small Ryugu samples available from *both* Chamber A (TD1) and Chamber C (TD2), whereas data from TD1 samples generally exhibit a larger degree of dispersion compared to TD2 samples (**Fig. 2**). Thus, the requisite mass of homogenized powder sample to obtain the $(X)_N$ value within ±5% error has been determined separately for TD1, TD2, and TD1+TD2, where the number of data is four or more (**Table 3**). Tantalum is not included in this table due to contamination from the projectile utilized for sampling, as discussed above (Nakamura *et al.*, 2022; Yokoyama *et al.*, 2023a). In the majority of cases, the requisite masses for TD1 samples exceed or are comparable to those of TD2 samples, and the requisite masses for TD1+TD2 are intermediate between those of TD1 and TD2. However, there are some exceptions. The requisite sample masses for Be and Bi are approximately two times greater for TD2 than TD1 samples, which may indicate the presence of nugget minerals incorporating these elements more in TD2 samples. Additionally, the requisite sample mass for Ne is considerably large for TD1+TD2, since only TD1 contains a few, solar wind-rich samples with exceptionally high Ne abundances (Okazaki *et al.*, 2023).

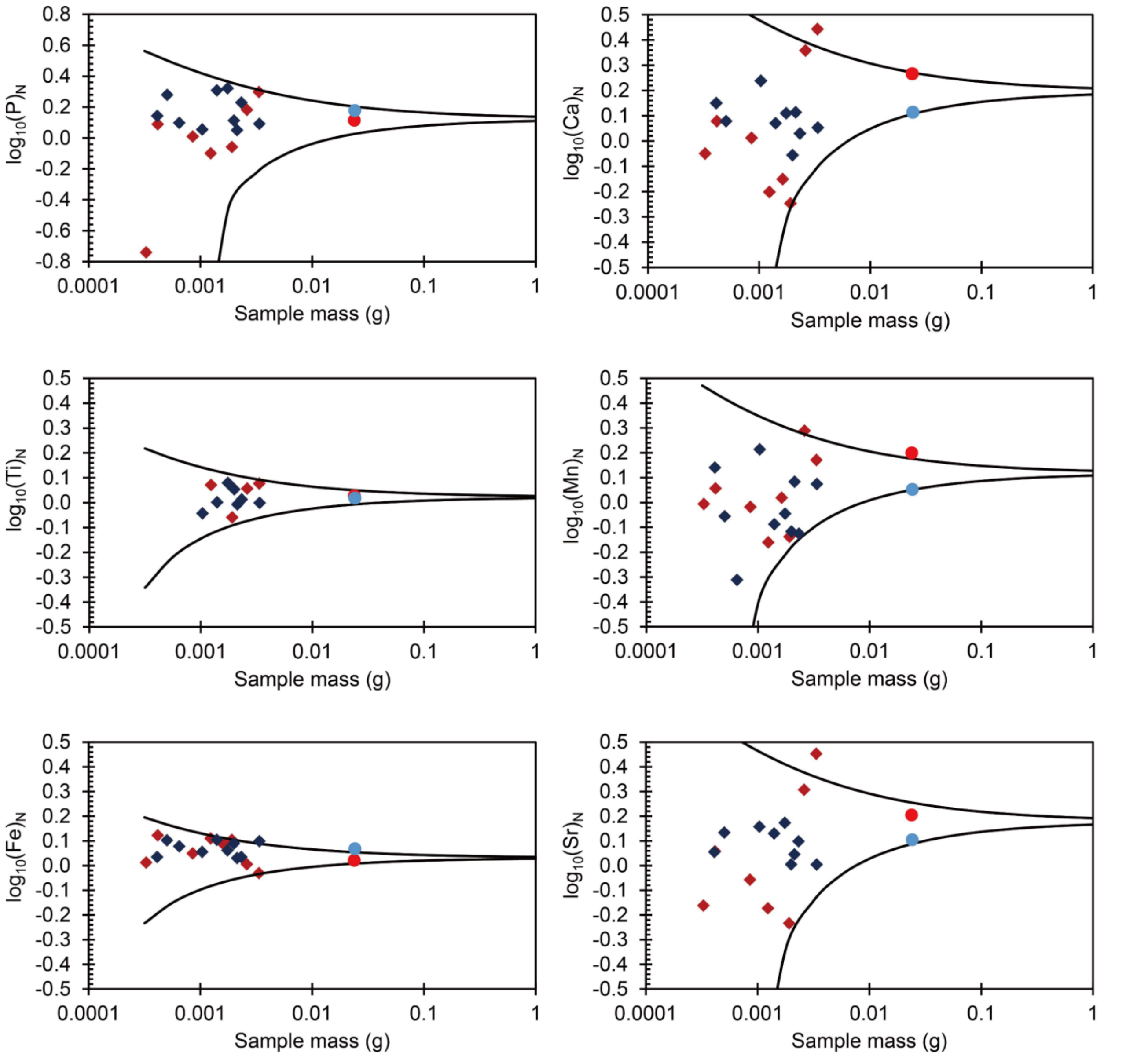

**Fig. 5** CI chondrite normalized elemental abundances for P, Ca, Ti, Mn, Fe, and Sr plotted against sample mass. Symbols are the same as in **Fig. 3**. Bold curves are the same as in **Fig. 4**.

**Figure 6** compares the requisite masses for TD1 and TD2 samples for the elements listed in **Table 3**, excluding highly volatile elements (H, C, and noble gases). To suppress the dispersion of $(X)_N$ values less than ±5%, approximately 750 and 400 mg of homogeneous powder samples are needed for TD1 (excluding B) and TD2 (excluding Be), respectively. The difference in the requisite mass between TD1 and TD2 samples is highlighted for P, Ca, Sr, and REEs, indicating that carbonates and phosphates are rather homogeneously distributed in TD2 compared to TD1 samples. The requisite mass can be reduced to 200 mg for TD1 and 100 mg for TD2 samples, if ±10% dispersion of $(X)_N$ values is allowed. In contrast, to obtain within ±3% error of $(X)_N$ values, 2000 and 1100 mg of homogenized powder samples are needed for TD1 and TD2, respectively. It should be noted that these requisite masses were calculated based on the limited number of data, mostly less than 10, for small Ryugu particles (< 3.3 mg). The acquisition of additional measurements of small Ryugu particles randomly obtained from each chamber will facilitate the prediction of the requisite mass to achieve a given dispersion of $(X)_N$ values.

**Table 3** Requisite mass for Ryugu samples to yield within ±5% variation for each element

| *Z* | *E* | TD1 (Chamber A) | | TD2 (Chamber C) | | TD1+TD2 | |
|---|---|---|---|---|---|---|---|
| | | mass (mg) | n# | mass (mg) | n# | mass (mg) | n# |
| 1 | H | 86 | 7 | 22 | 8 | 50 | 15 |
| 2 | He | 352 | 12 | 16 | 5 | 344 | 17 |
| 3 | Li | 55 | 7 | 31 | 9 | 40 | 16 |
| 4 | Be | 317 | 5 | 701 | 9 | 517 | 14 |
| 5 | B | 1307 | 4 | 200 | 7 | 726 | 11 |
| 6 | C | 178 | 8 | 162 | 10 | 156 | 18 |
| 10 | Ne | 1717 | 12 | 49 | 6 | 5390 | 18 |
| 11 | Na | 141 | 8 | 8 | 10 | 54 | 18 |
| 12 | Mg | 30 | 8 | 8 | 10 | 19 | 18 |
| 13 | Al | 44 | 8 | 8 | 10 | 24 | 18 |
| 14 | Si | 36 | 7 | 7 | 9 | 22 | 16 |
| 15 | P | 457 | 7 | 226 | 9 | 387 | 16 |
| 16 | S | 95 | 5 | 25 | 8 | 62 | 13 |
| 18 | Ar | 352 | 12 | 134 | 6 | 250 | 18 |
| 19 | K | 74 | 8 | 18 | 10 | 40 | 18 |
| 20 | Ca | 532 | 8 | 77 | 9 | 335 | 17 |
| 21 | Sc | 43 | 8 | 35 | 10 | 38 | 18 |
| 22 | Ti | 71 | 4 | 29 | 7 | 41 | 11 |
| 23 | V | 31 | 8 | 29 | 10 | 30 | 18 |
| 24 | Cr | 107 | 5 | 36 | 8 | 57 | 13 |
| 25 | Mn | 195 | 8 | 218 | 10 | 197 | 18 |
| 26 | Fe | 46 | 8 | 13 | 10 | 26 | 18 |
| 27 | Co | 65 | 8 | 9 | 10 | 30 | 18 |
| 28 | Ni | 54 | 8 | 29 | 10 | 38 | 18 |
| 29 | Cu | 226 | 7 | 272 | 9 | 238 | 16 |
| 30 | Zn | 43 | 8 | 21 | 10 | 29 | 18 |
| 31 | Ga | 141 | 8 | 104 | 10 | 114 | 18 |
| 33 | As | 117 | 4 | 51 | 4 | 129 | 8 |
| 34 | Se | 210 | 4 | 43 | 4 | 99 | 8 |
| 36 | Kr | 13 | 12 | 6 | 5 | 12 | 17 |
| 37 | Rb | 188 | 7 | 28 | 9 | 95 | 16 |
| 38 | Sr | 669 | 7 | 44 | 9 | 347 | 16 |
| 39 | Y | 727 | 7 | 10 | 9 | 394 | 16 |
| 40 | Zr | 107 | 4 | 27 | 7 | 48 | 11 |
| 41 | Nb | 32 | 4 | 5 | 7 | 13 | 11 |
| 42 | Mo | 85 | 4 | 25 | 7 | 40 | 11 |
| 48 | Cd | 46 | 7 | 8 | 9 | 23 | 16 |
| 50 | Sn | 173 | 4 | 37 | 7 | 108 | 11 |
| 51 | Sb | 359 | 5 | 146 | 8 | 196 | 13 |
| 54 | Xe | 14 | 14 | 4 | 6 | 10 | 20 |
| 55 | Cs | 181 | 7 | 200 | 9 | 199 | 16 |
| 56 | Ba | 223 | 7 | 63 | 9 | 139 | 16 |
| 57 | La | 276 | 7 | 17 | 9 | 138 | 16 |
| 58 | Ce | 307 | 7 | 20 | 9 | 163 | 16 |
| 59 | Pr | 335 | 7 | 19 | 9 | 168 | 16 |
| 60 | Nd | 328 | 7 | 27 | 9 | 183 | 16 |
| 62 | Sm | 446 | 7 | 27 | 9 | 227 | 16 |
| 63 | Eu | 601 | 7 | 27 | 9 | 315 | 16 |
| 64 | Gd | 627 | 7 | 14 | 9 | 332 | 16 |
| 65 | Tb | 607 | 7 | 23 | 9 | 316 | 16 |
| 66 | Dy | 701 | 7 | 15 | 9 | 360 | 16 |
| 67 | Ho | 701 | 7 | 21 | 9 | 368 | 16 |
| 68 | Er | 728 | 7 | 15 | 9 | 375 | 16 |
| 69 | Tm | 690 | 7 | 21 | 9 | 359 | 16 |
| 70 | Yb | 666 | 7 | 18 | 9 | 346 | 16 |
| 71 | Lu | 659 | 7 | 21 | 9 | 335 | 16 |
| 72 | Hf | 80 | 4 | 34 | 7 | 46 | 11 |
| 74 | W | 591 | 4 | 111 | 7 | 207 | 11 |
| 76 | Os | 124 | 4 | 25 | 4 | 64 | 8 |
| 77 | Ir | 415 | 4 | 167 | 4 | 261 | 8 |
| 81 | Tl | 126 | 7 | 118 | 9 | 115 | 16 |
| 82 | Pb | 37 | 7 | 14 | 9 | 23 | 16 |
| 83 | Bi | 250 | 7 | 388 | 9 | 312 | 16 |
| 90 | Th | 272 | 7 | 61 | 9 | 157 | 16 |
| 92 | U | 405 | 7 | 293 | 9 | 329 | 16 |

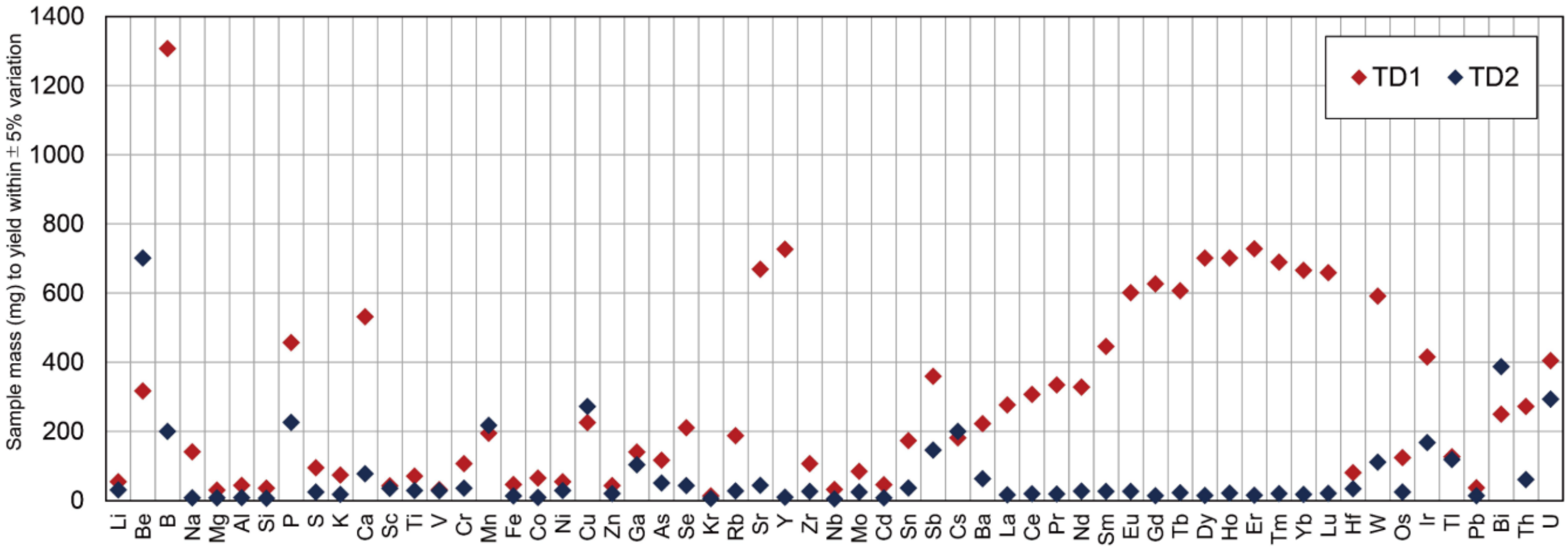

**Fig. 6** Requisite mass (mg) for TD1 and TD2 samples to yield within ±5% variation for the elements listed in **Table 3**, excluding highly volatile elements (H, C, and noble gases).

## 4. Future prospects of Ryugu sample analysis

The elemental abundances of CI chondrites have long been recognized as that representing the chemical composition of the Solar System - with some exceptions such as Li and highly volatile elements (Lodders, 2021; Lodders *et al.*, 2009; Palme and Zipfel, 2021; Palme *et al.*, 2014). On the other hand, there are discernible differences in refractory and other elemental abundances between Ryugu and CI chondrites (**Fig. 1**). Yokoyama et al. (2023a) posited that the general trend of the suprachondritic elemental abundances in Ryugu was possibly caused by the lower abundance of $H_2O$ in Ryugu (6.84 ± 0.34 wt.%) compared to CI chondrites (12.73 ± 0.63 wt.%). However, as shown in **Fig. 7**, the Fe-normalized elemental abundances for bulk Ryugu (Ryugu all) exhibit non-CI values exceeding the range of ±20% for some selected elements including Be, B, P, Ca, Mn, Sr, Nb, Pd, Ba, REEs (Eu–Lu), and Tl, while some other elements remain within the range of ±5% for both TD1 and TD2 samples (Al, S, Ti, Ni, Zn, Ga, As, Rb, Zr, Cd, Pb, and Th). This observation suggests that the low $H_2O$ abundance in Ryugu does not solely explain the observed differences between Ryugu and CI chondrite abundances. Such inconsistencies are particularly conspicuous for TD1 (Chamber A), while most of the Fe-normalized elemental abundances for TD2 (Chamber C) agree well with the CI chondrite data within the range of ±20%. In light of the distinctive, CI-like stable isotopic compositions observed for various elements in Ryugu (see Section 2.2), it is plausible that the primary building blocks of Ryugu and CI chondrites originated from a shared region in the early Solar System. Given the presence of some minor minerals enriched in specific elements involved in the Ryugu samples (see Section 3), the following three potential explanations can be put forward for the observed discrepancy in elemental abundances between Ryugu and CI chondrites:

1) CI-like Ryugu body: Both the TD1 and TD2 samples have CI-like chemical and isotopic compositions, but the Ryugu samples measured so far (< 80 mg) were biased in elemental abundances due to the nugget effect.
2) Partially CI-like Ryugu body: Either the TD1 or TD2, possibly TD2, has CI-like elemental abundances, while both have CI-like isotopic compositions.
3) Chemically non-CI Ryugu body: The asteroid Ryugu as a whole does not have CI-like elemental abundances, but has CI-like isotopic compositions.

To further investigate the chemical composition of Ryugu and its connection to CI chondrites, here we suggest establishing a new consortium to determine the representative elemental abundances of Ryugu by measuring aliquots from a large homogenized sample.

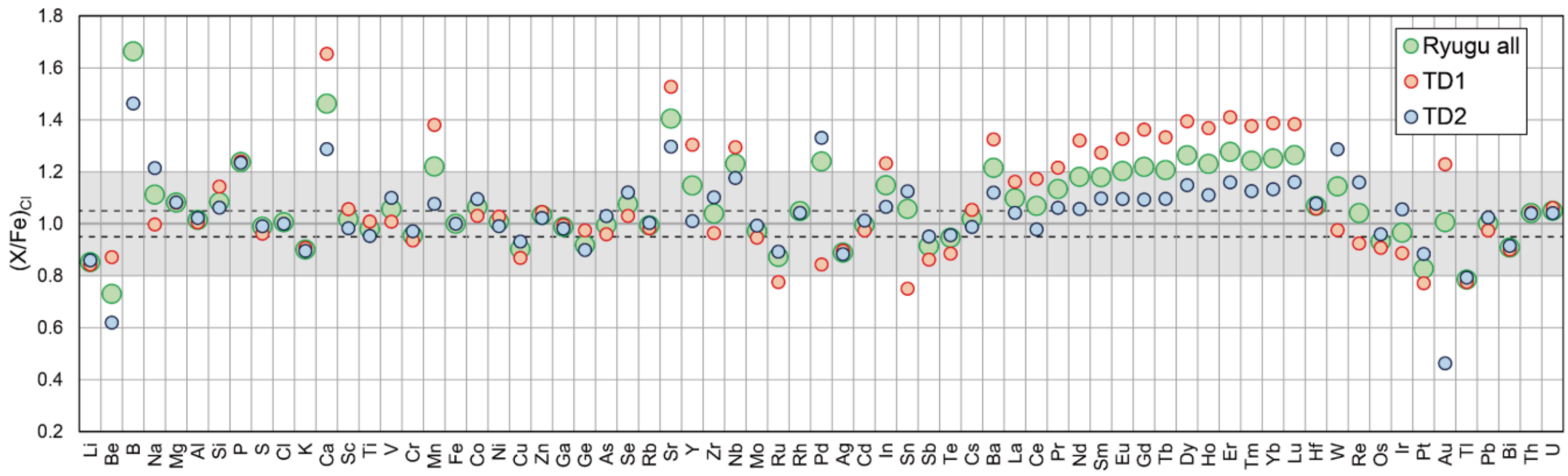

**Fig. 7** Fe- and CI-normalized elemental abundances in bulk Ryugu samples from Chambers A + C combined, Chamber A, and Chamber C. The dashed lines and gray shaded area indicate the ±5% and ±20% variation ranges, respectively.

As discussed above, the dispersion of the chemical composition and elemental ratios in published Ryugu data (**Figs. 3-4**) are attributable to non-representative sampling of mineral phases highly enriched in some elements, which are likely carbonates and phosphates. This nugget effect can be mitigated by measuring homogenized material prepared from a relatively large sample mass. Considering the uncertainties associated with XRF and ICP-MS measurements for Ryugu (< 3% for major elements, < 5% for trace elements: Yokoyama *et al.*, 2023a), analysis using a powder sample exhibiting within ±5% elemental heterogeneity is an achievable goal. Excluding Be, B, and Bi, this necessitates the use of at least 750 mg and 400 mg of homogenized TD1 and TD2 samples, respectively (**Fig. 6**). The stored Ryugu samples suitable for the preparation of the powdered Ryugu would be an aggregate of many submillimeter grains from each Chambers A (TD1) and C (TD2). Combining samples from different chambers to produce a homogeneous Ryugu powder, collected from the two different touch-down sites on the Ryugu body, should be avoided. The mixing of samples from the two chambers will impede the ability to investigate the chemical heterogeneity present in the Ryugu samples from distinct locations. The preparation of two distinct powders derived from the two chambers is essential for testing the three aforementioned scenarios and for understanding the evolutionary history of the Ryugu parent body, including the effect of aqueous alteration and incorporation of elements from the solar wind.

The ultimate goal of such a consortium is to provide new insights into the chemical composition of the Solar System. The precise estimation of the Solar System composition using currently all available Ryugu elemental abundances is challenging due to the nugget effect. Therefore, it is of great importance to measure homogeneous Ryugu powder samples prepared from a statistically significant sample mass. In consideration of the statistical analysis presented above and the total sample mass collected in each chamber (A: 3.237±0.003 g, C: 2.025±0.003 g, Yada *et al.*, 2022), it is recommended that the sample mass homogenized for the consortium be 1 g (at least 0.75 g) for Chamber A and 0.5 g (at least 0.4 g) for Chamber C. Furthermore, within the context of the consortium, it is strongly advised that the elemental abundances of CI chondrites (if possible, Orgueil, Ivuna, and Alais) be simultaneously measured using the identical analytical techniques as those employed for Ryugu. These CI chondrite samples should be prepared by pulverizing a mass similar to that of Ryugu. This will result in updating and improvement of the reference values for CI chondrites.

The newly obtained, more precise chemical composition of Ryugu and CI chondrites will be used by multidisciplinary communities in various scientific fields including astronomy, astrobiology, cosmochemistry, geochemistry, geology, and planetary physics. For instance, comparative studies of the chemical compositions of chondrites, Ryugu, and the terrestrial mantle will facilitate the discussion on the origin of Earth and rocky planets, including the origin of Earth's water. Another comparative study of elemental abundances between CI chondrites, Ryugu, and new asteroidal materials collected from B-type asteroid (101955) Bennu by the OSIRIS-REx mission (Lauretta *et al.*, 2024) will expand our knowledge regarding the formation of primordial small bodies in the early Solar System, and the chemical heterogeneity of the solar nebula. More importantly, the consistency must be evaluated between the new Ryugu and CI data, as well as the elemental abundances of the solar photosphere determined by improved spectroscopic measurements coupled with sophisticated atmospheric models. The revised Solar System composition will facilitate comprehension of elemental abundances in other astronomical objects including stars and the interstellar medium. Furthermore, it will serve as a fundamental anchor point at 4.5 Ga for evaluating various Galactic Chemical Evolution models.

**Acknowledgments** We thank H. Yurimoto and all the members of the Hayabusa2 Initial Analysis Chemistry Team and the Initial Analysis Core. Discussion with team members of the Hayabusa2 Sample Allocation Committee (HSAC) has improved the quality of this paper. TY thanks to F. Wombacher for his comments on the elemental abundance analysis of Ryugu. We are also indebted to K. Terada, K. Lodders, and an anonymous reviewer for their editorial assistance and constructive feedback. This research was supported by JSPS KAKENHI Grants to TY (23H00143, 21KK0058, and 20H04609).